\documentclass[11pt]{article} 
\advance\voffset by 1.2truecm
\advance\hoffset by -.5truecm
\usepackage{fullpage,subeqn}
\usepackage{epsf,graphicx}
\def\be{\begin{equation}}
\def\ee{\end{equation}}
\def\beq{\begin{eqnarray}}
\def\eeq{\end{eqnarray}}

\def\gsim{\lower0.5ex\hbox{$\:\buildrel >\over\sim\:$}}
\def\lsim{\lower0.5ex\hbox{$\:\buildrel <\over\sim\:$}}

\begin{document}
\thispagestyle{empty}

\vspace{2em}

\begin{center}
\Large{\bf NEUTRINO SYMMETRIES FROM HIGH TO LOW SCALES} \\[1cm]
\large{Probir Roy} \\[.6cm]
{\sl Tata Institute of Fundamental Research,} \\ 
{\sl Homi Bhabha Road, Mumbai 400 005, INDIA} \\ 
\end{center}

\vspace{2em}
\pretolerance=10000

Proposed symmetry relations, e.g., quark-lepton complementarity (QLC)
or tribimaximal mixing (TBM), need to be imposed at a high scale
$\wedge \sim 10^{12}$ GeV characterising the large masses of
right-handed neutrinos required to implement the seesaw mechanism. RG
evolution down to the laboratory scale $\lambda \sim 10^3$ GeV,
generically prone to spoil these relations and their predicted neutrino
mixing patterns, can be made to preserve them by appropriately
constraining the Majorana phases $\alpha_{2,3}$. This is explicitly
demonstrated in the MSSM for two versions of QLC and two versions of
TBM. A preference for $\alpha_2 \simeq \pi$ (i.e. $m_1 \simeq - m_2$)
emerges in each case. Discrimination among the four cases is shown to
be possible by future measurements of $\theta_{13}$. 

\vspace{3em}

\noindent{\bf Preliminaries}\\

The unitary neutrino mixing matrix $U^\nu$ acts between mass
eigenstates $|i> (i = 1,2,3)$ and flavour eigenstates $|a> (a = e,
\mu, \tau)$ by \cite{one}
\be
|a> = (U^\nu)_{ai} |i> .
\ee
The SM fermion mass term in the chiral flavour basis is 
\be
{\cal L}_m = - \bar f_{La} M^f_{ab} f_{Rb} + h.c.
\ee
Let the unitary transformations of $f_{La}$ and $f_{Rb}$ to the mass
basis be 
\begin{subequations} 
\beq
f_{La} &=& U^f_{ai} f_{Li}, \\ [3mm] 
f_{Rb} &=& W^f_{bj} f_{Rj} . 
\eeq
\end{subequations} 
Thus, in the mass basis, 
\be
{\cal L}_m = - \bar f_{Li} {\cal M}^{f(D)}_{ij} f_{Rj} + h.c. ,
\ee
where the diagonal mass matrix ${\cal M}^{f(D)}$ is given by the
biunitarily transformed $M^f$: 
\be 
{\cal M}^{f(D)} = U^{f^\dagger} M^f W^f ,
\ee

The charged current  weak interaction for quarks is 
\begin{subequations}
\beq
{\cal L}_{cc} &=& \bar u_{La} \gamma^\mu d_{La} W^+_\mu + h.c. =
\bar u_{Li} \gamma^\mu V^{CKM}_{ij} d_{Lj} W^+_\mu + h.c., \\ [3mm]
V^{CKM} &=& U^{u^\dagger} U^d . 
\eeq
\end{subequations}
The corresponding interaction in the lepton sector can be written as
\begin{subequations}
\beq
{\cal L}^\ell_{cc} &=& \bar\ell_{La} \gamma^\mu \nu_{La} W^-_\mu +
h.c. = \bar\ell_{Li} \gamma^\mu U^{PMNS}_{ij} \nu_{Lj} W^-_\mu +
h.c. , \\ [3mm]
U^{PMNS} &=& U^{\ell^\dagger} U^\nu . 
\eeq
\end{subequations}

The above similarity leads one to attribute a CKM-like form \cite{one}
to $U^{PMNS}$ in terms of three angles $\theta_{12}, \theta_{23},
\theta_{13}$ and $a$ Dirac phase $\delta_\ell$: 
\be
U^{CKM-form} = \left(\matrix{
c_{12}c_{13} & s_{12}c_{13} & s_{13}e^{-i\delta_\ell} \cr
-s_{12}c_{23}-c_{12}s_{23}s_{13}e^{i\delta_\ell} &
c_{12}c_{23}-s_{12}s_{23}s_{13}e^{i\delta_\ell} & s_{23}c_{13} \cr 
s_{12}s_{23}-c_{12}c_{23}s_{13}e^{i\delta_\ell} & -c_{12}s_{13}-s_{12}
c_{23}s_{13}e^{i\delta_\ell} & c_{23}c_{13}}\right),  
\ee
with $c_{ij} = \cos\theta_{ij}$ and $s_{ij} =
\sin\theta_{ij}$. However, for Majorana neutrinos, there is an
additional diagonal matrix factor containing two more phases (since
one can be absorbed in the overall neutrino phase): 
\be
U^{PMNS} = U^{CKM-form} diag. (1, e^{-i{\alpha_2 \over 2}},
e^{-i{\alpha_3 \over 3}}). 
\ee
Eq.(9) means that, for Majorana neutrinos with a mass term
\be
{\cal L}_m = - {1\over 2} \bar\nu^c {\cal M}^\nu \nu + h.c. ,
\ee
(5) in fact becomes
\be
{\cal M}^{\nu(D)} = U^{\nu^\dagger} {\cal M}^\nu U^{\nu^\ast} =
diag. (m_1, m_2, m_3).
\ee
Given (9), one can take $m_1 = |m_1|, m_2 = |m_2| e^{i\alpha_2}, m_3 =
|m_3| e^{i\alpha_3}$ in (11). 

\bigskip

\noindent{\bf Neutrino factfile} \\

We now know that at least two of the three light neutrinos are
massive. In the notation of (11), one can already make 
statements \cite{two} on three scales respectively, namely solar, atmospheric and
cosmological: 
\begin{subequations}
\beq
\sqrt{\delta m_S^2} &\equiv& [|m_2|^2 - |m_1|^2]^{1/2} \sim 0.009 \ eV,\\
\sqrt{\delta m_A^2} &\equiv& \big| |m_3|^2 - |m_2|^2\big|^{1/2} \sim
0.05 \ eV,\\
\sum^3_{i=1} |m_i| & < & {\cal O} (1) \ eV.
\eeq
\end{subequations}

Furthermore, while the value of the Dirac phase $\delta_\ell$ in (8)
is unknown, we do know what two of the angles in $U^{PMNS}$ are and
have an upper bound on the third: 
\be
\theta_{12} = 33.8^{\circ + 2.4^\circ}_{-1.8^\circ}, \ \theta_{23} =
45^\circ \pm 4^\circ , \ \theta_{13} < 13^\circ, \delta_\ell = ? 
\ee
The magnitudes of the elements of the matrices $V^{CKM}$ and
$U^{PMNS}$ are now roughly known to be 
\be
|V^{CKM}| = \left(\matrix{0.97 & 0.22 & 0.003 \cr 0.22 & 0.97 & 0.04
\cr 0.01 & 0.04 & 0.99} \right), \ |U^{PMNS}| \sim \left( \matrix{
0.8 & 0.5 & < 0.14 \cr 0.4 & 0.6 & 0.7 \cr 0.4 & 0.6 &
0.7}\right) 
\ee
and present a striking contrast between small and large deviations
from the unit matrix in the two cases respectively. 

We don't yet know if the ordering of the neutrino masses is `normal'
$(|m_3| > |m_2| > |m_1|)$ or `inverted' $(|m_2| > |m_1| > |m_3|)$. So
three mass patterns for the three neutrinos are still possible: 
\begin{enumerate}
\item[{(1)}] Normal hierarchical (NH): $|m_1| \ll |m_2| \sim 0.009 \ eV
\ll |m_3| \sim 0.05 \ eV$, 
\item[{(2)}] Inverted hierarchical (IH): $|m_3| \ll |m_1| \lsim |m_2|
\sim 0.05 \ eV$, 
\item[{(3)}] Quasi-degenerate (QD): $0.05 eV < |m_1| \sim |m_2| \sim
|m_3| \lsim {\cal O} (0.33) \ eV$. 
\end{enumerate}
Within the QD pattern, the mass ordering could be either normal or
inverted. We generically club the IH and QD cases under the heading
`nonhierarchical' (NH): 
\[
NH \equiv \{ IH, QD\}.
\]

\bigskip

\noindent{\bf Mass parametrization with Majorana phases} \\

We introduce three real parameters, two of them dimensionless
($\rho_A$ and $\epsilon_S$) and one dimensional $(m_0)$, such that 
\begin{subequations}
\beq
|m_1| &=& m_0 (1 - \rho_A) (1 - \epsilon_S) ,  \\
|m_2| &=& m_0 (1 - \rho_A) (1 + \epsilon_S) ,  \\
|m_3| &=& m_0 (1 + \rho_A) , 
\eeq
\end{subequations}
with $0 < \epsilon_S \leq 1, -1 \leq \rho_A \leq 1, 0 < m_0 < {\cal O}
(0.33)$ eV. The sign of $\rho_A$ is positive (negative) for a normal
(inverted) mass ordering. The solar as well as atmospheric mass scales
and the sum of the neutrino masses are given respectively by 
\begin{subequations}
\beq
\delta m^2_S &=& 4 m^2_0 (1 - \rho_A)^2 \epsilon_S \\
|\delta m^2_A| &=& 4 m^2_0 |\rho_A| , \\
\sum_i |m_i| &=& 3m_0 \left(1 - {\rho_A \over 3}\right). 
\eeq
\end{subequations}

It is convenient to define a derived dimensionless parameter $\Gamma$
by
\be
\Gamma \equiv \rho^{-1}_A - \rho_A
\ee
which is positive (negative) for a normal (inverted) mass ordering and
is allowed by the present data to be anywhere between zero and $\pm$
182. {\it Sample values} of these quantities are given in Table 1 for
the three mass patterns. 
\begin{table}[h!]
\begin{center}
\begin{tabular}{|lcl|}\hline
NH & : & $m_0 \simeq 0.025 \ eV, \rho_A \simeq 1, \epsilon_S \simeq 1,
\Gamma = 0 +$ \\
IH & : & $m_0 \simeq 0.025 \ eV, \rho_A \simeq -1, \epsilon_S \simeq 1.6
\times 10^{-2}, \Gamma = 0 -$ \\
QD & : & $0.025 \ eV \ll m_0 \lsim 0.33 \ eV, |\rho_A| \lsim 0.056, \epsilon_S \gsim 2
\times 10^{-6}, |\Gamma| \lsim 182$. \\ \hline
\end{tabular}
\caption{Sample values of $m_0, \rho_A, \epsilon_S$ and $\Gamma$ for
the three mass patterms.} 
\end{center}
\end{table}

\newpage

\noindent{\bf Running neutrino masses and mixing angles} \\

Loop divergences and corresponding renormalization procedures turn 
coupling strengths $g_i$ into functions of the evolution variable $t = (16
\pi^2)^{-1} ln \ Q/\wedge$, where $Q$ is the running energy and
$\wedge$ some fixed (high) scale. In particular, this is true of the
fermionic Yukawa couplings relevant to neutrino masses and mixing
angles. As a result, the latter become functions of $t : m_i \to
m_i(t), \theta_{ij} \to \theta_{ij} (t), \delta_\ell \to \delta_\ell
(t), \alpha_{2,3} \to \alpha_{2,3} (t)$. 

Our basic idea \cite{three} is to consider certain neutrino
symmetries, which fix the neutrino mixing pattern, to be operative at
a high scale $Q=\wedge$. We choose $\wedge \sim 10^{12}$ GeV
characterising the mass scale of heavy right handed neutrinos
responsible for the seesaw origin of tiny neutrino masses. We then see
the effects of RG evolution \cite{four} down to a laboratory scale
$\lambda \sim 1$ TeV on that pattern. We do so within the Minimal
Supersymmetric Standard Model (MSSM) \cite{five} which is why we have
chosen $\lambda$ to be of the order of the expected scale of soft
supersymmetry breaking. While one can debate the precise values of
$\wedge$ and $\lambda$ that have been chosen,  the effects that we are
concerned with are only logarithmically sensitive to
$\wedge/\lambda$. Moreover, the RG effects are controlled by factors
such as $|m_i + m_j|^2 (|m_i|^2 - |m_j|^2)^{-1} \Delta_\tau$ where the
dimensionless parameter $\Delta_\tau \lsim {\cal O} (10^{-2})$. For neutrinos
with a normal hierarchy, this is negligible. Only for nonhierarchical
neutrinos are these effects significant. 

The neutrino Majorana mass matrix originates at the scale $\wedge$
from a dimension 5 operator
\be
{\cal O} = c_{\alpha\beta} {(\ell_\alpha \cdot H) (\ell_\beta \cdot H)
\over \wedge} .
\ee
In (18), $\ell_\alpha$ and $H$ are the $SU(2)$ doublet lepton and
Higgs fields respectively, with $\alpha$ being a generation index and
$c_{\alpha \beta}$ being dimensionless coefficients that run with the
energy scale. Then 
\be
({\cal M}^{\nu,\wedge})_{\alpha\beta} \sim c_{\alpha\beta} {v^2 \over
\wedge} , 
\ee
with $v$ = 246 GeV and $\wedge \sim M_{MAJ}$, the Majorana mass
characterising the set of heavy SU(2)-singlet Majorana neutrinos
$\{N\}$. The coefficients $c_{\alpha \beta}$ are the ones which evolve
from $Q = \wedge$ to $Q = \lambda$. 

One-loop contributions to the evolution of ${\cal M}^\nu$ from gauge
bosons, gauginos and sfermions of the MSSM lead to the relation
\cite{six} 
\be
{\cal M}^{\nu,\lambda} = I_k I^T_\kappa {\cal M}^{\nu,\wedge} I_\kappa
\ee
with
\begin{subequations}
\beq
I_k &=& \exp \left[ - \int^{t(\lambda)}_0 d\tau \{ 6 g^2_2 (\tau) + 2
g^2_Y (\tau) - 6 Tr (Y^\dagger_u Y_u) (\tau)\} \right] , \\
I_\kappa &=& \exp \left[ - \int^{t(\lambda)}_0 d\tau (Y^\dagger_\ell
Y_\ell) (\tau)\right] . 
\eeq
\end{subequations}
In (21), $g_{2,Y}$ are the SU(2), U(1) gauge couplings and
$Y_{u,\ell}$ are the Yukawa coupling matrices in family space for
up-quarks, charged leptons. Let us define a dimensionless quantity 
\be
\Delta_\tau \equiv m^2_\tau (\tan^2 \beta + 1) (8 \pi^2 v^2)^{-1} ln 
{\wedge \over \lambda} , 
\ee
where $\tan\beta = v_u/v_d, v_{u,d}$ being the VEV of the neutral
Higgs coupling to up, down type of fermions. This $\Delta_\tau \lsim
10^{-2}$ for $\tan \beta \lsim 30$. 

In the basis with $U^\ell = I$, i.e., ${\cal M}^\ell = diag. (m_e,
m_\mu, m_\tau)$, to linear order in $\Delta_\tau$ one has \cite{three}
\be
I_\kappa \simeq diag. (1, 1, 1 - \Delta_\tau) + {\cal O}
(\Delta^2_\tau) .
\ee
Now (20) can be rewritten as
\be
{\cal M}^{\nu,\lambda} = I_k diag. (1, 1, 1 - \Delta_\tau) {\cal
M}^{\nu,\lambda} diag. (1, 1, 1 - \Delta_\tau) + {\cal O}
(\Delta^2_\tau) . 
\ee
Since in this basis the unitary matrix $U^\nu$ diagonalising ${\cal
M}^\nu$ equals $U^{PMNS}$, we can use (24) to relate $U^{PMNS,\wedge}$
and $U^{PMNS,\lambda}$ and consequently $m^\wedge_i,
\theta^\wedge_{ij}, \delta^\wedge_\ell$ to $m^\lambda_i,
\theta^\lambda_{ij}, \delta^\lambda_\ell$ respectively. Though we make
these approximations to facilitate the use of analytically transparent
expressions, our final results, shown in later figures, are based on
the numerical integration of the full equations of Antusch et
al. \cite{four}. 

The high scale symmetries considered by us dictate
$\theta^\wedge_{13}$ to be a small parameter which is
negligible. Moreover, $\theta^\wedge_{13} \Delta_\tau < 10$ and so
${\cal O} (\theta^\wedge_{13} \Delta_\tau)$ terms can also be
neglected. Then the evolution of all the above parameters can be
computed analytically in a simple manner. We have 
\begin{subequations}
\beq
\theta^\lambda_{12} &=& \theta^\wedge_{12} + k_{12} \Delta_\tau +
{\cal O} (\theta^\wedge_{13} \Delta_\tau, \Delta^2_\tau) , \\ [1mm]
\theta^\lambda_{23} &=& \theta^\wedge_{23} + k_{23} \Delta_\tau +
{\cal O} (\theta^\wedge_{13} \Delta_\tau, \Delta^2_\tau) , \\ [1mm]
\theta^\lambda_{13} &=& k_{13} \Delta_\tau + {\cal O}
(\theta^\wedge_{13}, \Delta^2_\tau) , \\ [1mm] 
\alpha^\lambda_{2,3} &=& \alpha^\wedge_{2,3} + a_{2,3} \Delta_\tau +
{\cal O} (\theta^\wedge_{13} \Delta_\tau, \Delta^2_\tau) , \\ [1mm]
|m^\lambda_i| &=& I_k |m^\wedge_i| \left[1 + \mu_i \Delta_\tau + {\cal O}
(\theta^\wedge_{13} \Delta_\tau, \Delta^2_\tau)\right], \\ [1mm]
\delta^\ell &=& d_\ell \Delta_\tau + {\cal O} (\theta^\wedge_{13}
\Delta_\tau + \Delta^2_\tau) . 
\eeq
\end{subequations}
The values of $k_{ij}$ are
\begin{subequations}
\beq
k_{12} &=& {1\over 2} \sin 2 \theta^\wedge_{12} \sin^2
\theta^\wedge_{23} {|m^\wedge_1 + m^\wedge_2|^2 \over |m^\wedge_2|^2 -
|m^\wedge_2|^2} \nonumber \\ [1mm]
&\simeq& {1\over 4 \epsilon^\wedge_S} \sin 2
\theta^\wedge_{12} \sin^2 \theta^\wedge_{23} 
\left[ 1 + \cos \alpha^\wedge_2 + \epsilon^{\wedge^2}_S (1 - \cos
\alpha^\wedge_2)\right] + {\cal O} (\theta^\wedge_{13}), \\ [1mm]  
k_{23} &=& {1\over 2} \sin 2 \theta^\wedge_{23} \left( \cos^2
\theta^\wedge_{12} {|m^\wedge_1 + m^\wedge_3|^2 \over |m^\wedge_3|^2 -
|m^\wedge_2|^2} + \sin^2 \theta^\wedge_{12} {|m^\wedge_1 +
m^\wedge_2|^2 \over |m^\wedge_2|^2 - |m^\wedge_1|^2} \right)
\nonumber\\ [1mm]
&\simeq& {\Gamma^\wedge \over 4} \sin 2 \theta^\wedge_{23}  
\Big[1 + \cos^2\theta^\wedge_{12} \cos (\alpha^\wedge_2 -
\alpha^\wedge_3) \nonumber \\ [1mm] 
&& +
\sin^2 \theta^\wedge_{12} \cos \alpha^\wedge_3 \Big] + {\rho_A \over
2} \sin 2 \theta^\wedge_{12} \sin 2 \theta^\wedge_{23} + {\cal O}
(\theta^\wedge_{13}, \epsilon^\wedge_S), \\ [1mm]
k_{13} &=& {1\over 4} \sin 2 \theta^\wedge_{12} \sin 2
\theta^\wedge_{23} \Big({|m^\wedge_2 + m^\wedge_3|^2 \over
|m^\wedge_3|^2 - |m^\wedge_2|^2} - {|m^\wedge_1 + m^\wedge_2|^2 \over
|m^\wedge_2|^2 - |m^\wedge_1|^2} \nonumber \\ [1mm] 
&\simeq&  
{\Gamma^\wedge \over 8} \sin 2 \theta^\wedge_{12} \sin 2
\theta^\wedge_{23} 
\left[\cos (\alpha^\wedge_2 - \alpha^\wedge_3) -
\cos \alpha^\wedge_3 \right] + {\cal O}
(\epsilon^\wedge_S, \theta^\wedge_{13}\Big). 
\eeq
\end{subequations}
The sign of $k_{12}$ is {\it always} positive, as is clear from
(26a). 

\bigskip

\noindent{\bf High scale neutrino symmetries} \\

We consider four cases in this category: QLC 1 \cite{seven}, QLC 2
\cite{seven,eight}, TBM 1 \cite{nine} and TBM 2 \cite{ten}. None of
these determines the ordering of the neutrino masses which can be
normal or inverted in each case. Recall first the respective forms of
$U$ for bimaximal and tribimaxial mixing. 
\begin{subequations}
\beq
U^{BM} &=& \left(\matrix{1/\sqrt{2} & 1/\sqrt{2} & 0 \cr -1/2 & 1/2 &
1/\sqrt{2} \cr 1/2 & - 1/2 & 1/\sqrt{2}} \right), \theta^\wedge_{23} =
\pi/4 = \theta^\wedge_{12}, \ \theta^\wedge_{13} = 0 , \\ 
U^{TBM} &=& {1 \over \sqrt{6}} \left(\matrix{2 & \sqrt{2} & 0 \cr -1
& \sqrt{2} & \sqrt{3} \cr -1 & - \sqrt{2} & \sqrt{3}}\right),
\theta^\wedge_{23} = \pi/4, \ \theta^\wedge_{12} = \sin^{-1} {1\over
\sqrt{3}} \simeq 35.26^\circ, \ \theta^\wedge_{13} = 0 . 
\eeq
\end{subequations}
In (27) we have also listed the corresponding values of the mixing
angles at the scale $\wedge$ where these symmetries are
implemented. The content of that implementation in each case is
summarised in Table 2. 

\begin{table}[h!]
\scriptsize{
\begin{center}
\begin{tabular}{|l|l|l|l|} \hline
\qquad\qquad QLC 1 \qquad\qquad & \qquad\qquad QLC 2 \qquad\qquad & \qquad\qquad TBM 1 \qquad\qquad &
\qquad\qquad TBM 2 \qquad\qquad \\ \hline
$U^{PMNS} = V^{CKM^\dagger} U^{\nu,BM}$ & $U^{PMNS} =
U^{\ell,BM^\dagger} V^{CKM}$ & $U^{PMNS} = U^{\nu,TBM}$ & $U^{PMNS} =
\tilde V^{CKM^\dagger} U^{\nu,TBM}$ \\
$U^u=I$ basis $\to U^d = U^\ell$ & $U^d=I$ basis $\to U^u = U^\nu$ &
Family $A_4$ or $S_3$ & $\tilde V^{CKM} \simeq \left(\matrix{1 &
\theta_{C/3} & 0 \cr \theta_{C/3} & 1 & -|V_{cb}| \cr 0 & |V_{cb}| & 1}
\right) + {\cal O} (\theta^3_C)$ \\
$\theta^\wedge_{12} \simeq {\pi \over 4} - \theta_C/\sqrt{2} \simeq 
35.4^\circ$ & $\theta^\wedge_{12} \simeq {\pi \over 4} - \theta_C \simeq
32.4^\circ$ & $\theta^\wedge_{12} = \sin^{-1} {1 \over \sqrt{3}} \simeq
35.3^\circ$ & $\theta^\wedge_{12} \simeq \sin^{-1} {1 \over \sqrt{3}} -
{\theta_C \over 3\sqrt{2}} \simeq 32.3^\circ$ \\ 
$\theta^\wedge_{23} \simeq {\pi \over 4} - |V_{cb}| - {\theta^2_C
\over 4} \simeq 42.1^\circ$ & $\theta^\wedge_{23} \simeq {\pi \over 4}
- {|V_{cb}| \over \sqrt{2}} \simeq 43.4^\circ$ & $\theta^\wedge_{23} =
45^\circ$ & $\theta^\wedge_{23} \simeq {\pi \over 4} - |V_{cb}| \simeq
42.7^\circ$ \\ 
$\theta^\wedge_{13} \simeq {\theta_C \over \sqrt{2}} \simeq 8.9^\circ$
& $\theta^\wedge_{13} \simeq {|V_{cb}| \over \sqrt{2}} \simeq 1.6^\circ$
& $\theta^\wedge_{13} = 0$ & $\theta^\wedge_{13} \simeq {\theta_C
\over 3 \sqrt{2}} \simeq 3.1^\circ$ \\ \hline
\end{tabular}
\caption{Statement and consequence of each high scale symmetry}
\end{center}}
\end{table}

\noindent{\bf Correlated constraints} \\

The 3$\sigma$ allowed ranges \cite{one} for neutrino mass and mixing
parameters are tabulated below. 
\begin{table}[h!]
\begin{center}
\begin{tabular}{|c|} \hline
$7 \times 10^{-5} eV^2 < \delta m^2_S < 9.1 \times 10^{-5} eV^2$ \\
$1.7 \times 10^{-3} eV^2 < |\delta m^2_A| < 3.3 \times 10^{-3} eV^2$\\
$30^\circ < \theta_{12} < 39.2^\circ$ \\
$35.5^\circ < \theta_{23} < 55.5^\circ$ \\ 
$\theta_{13} < 12^\circ$ \\  \hline
\end{tabular}
\caption{3$\sigma$ allowed values of neutrino mass and mixing
parameters} 
\end{center}
\end{table}

\newpage

\noindent The tightest constraints come from $\theta_{12}$. Figure 1 shows
exclusion regions in the $m_0^\wedge \tan\beta - \alpha^\wedge_2$
plane for each high scale symmetry considered. The peak at $\alpha_2
\simeq \pi$ shows a preference in all these models for the approximate
result $m_1 \simeq - m_2$  which is also a desired result for
leptogenesis \cite{eleven} with nonhierarchical neutrinos. 
\begin{figure}[h!]
\begin{center}
\vspace*{2in}
\includegraphics{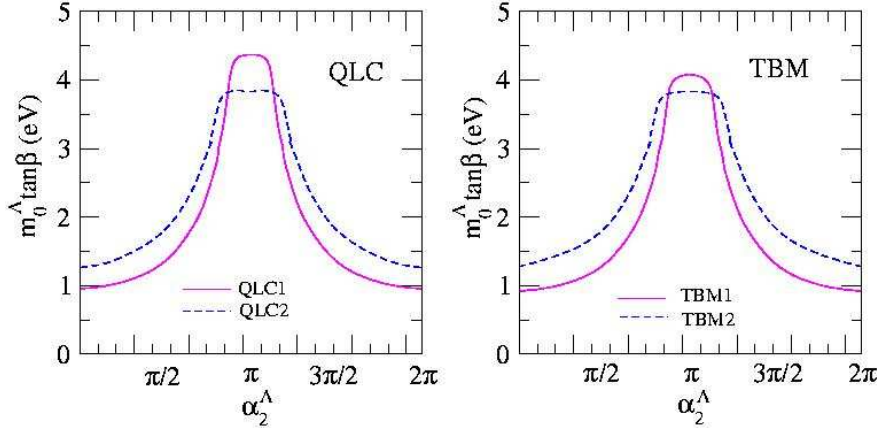}
\end{center}
\caption{Exclusion regions (above the curves) in the $m^\wedge_0
\tan\beta - \alpha^\wedge_2$ plane.}
\end{figure}
The positivity of $k_{12}$ dictates that the measured value of
$\theta_{12}$ should {\it exceed} 35.4$^\circ$, 32.4$^\circ$,
35.26$^\circ$ and 35.3$^\circ$ for the QLC 1, QLC 2, TBM 1 and TBM 2
cases respectively. In particular, reduced errors on $\theta_{12}$,
possibly from SNO 3, can really put QLC 1 and TBM 2 out of commission. 

Turning to the other mixing angles, the measured value of
$\theta_{23}$ has to exceed (be less than) 42.5$^\circ$, 42.7$^\circ$,
45$^\circ$, 42.5$^\circ$ for the QLC 1, QLC 2, TBM 1 and TBM 2 cases
respectively for a normal (inverted) ordering of neutrino masses. On
the other hand, $\Delta \theta_{13} = \theta^\wedge_{13} -
\theta^\lambda_{13}$ depends on $m_0 \tan \beta$. The allowed regions
in the $\theta^\lambda_{13} - m^\wedge_0 \tan \beta$ plane for the
four cases are shown in Fig. 2. In particular, a measured value of
$\theta_{13} < 6^\circ$ will exclude QLC 1. Also, if $m^\wedge_0 \tan
\beta < 2$ eV (i.e. $m^\lambda_0 \tan \beta < 1.4$ eV), TBM 2 will be
distinguishable from QLC 2 and TBM 1. Finally, values of $m^\wedge_0
\tan \beta > 4.4$ eV (i.e. $m^\lambda_0 \tan \beta > 3.1$ eV) are
disallowed for all four cases. 
\begin{figure}[h!]
\begin{center}
\vspace*{2in}
\includegraphics{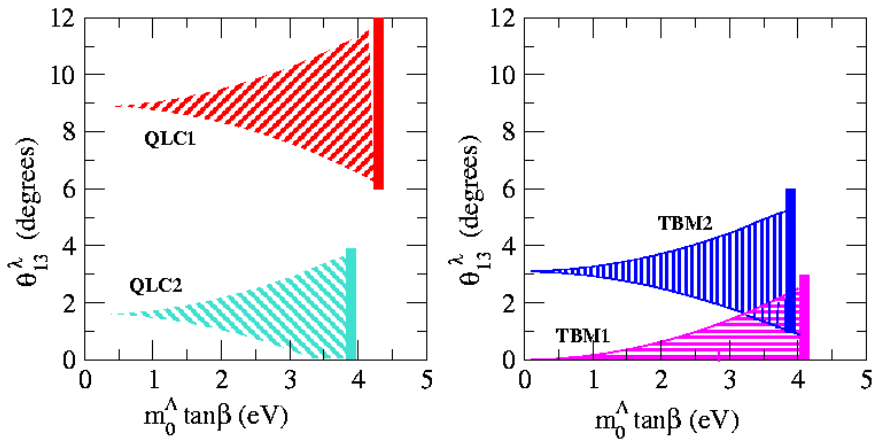}
\end{center}
\caption{Allowed regions in the $\theta_{13} - m^\wedge_0
\tan\beta$ plane for all four cases.}
\end{figure}

\eject

This talk is based on the work reported in Ref.[3]. The author
acknowledges the hospitality of the Te\'orica de Fisica de Particulas,
Instituto Superior T\'ecnico, Lisbon.

\end{document}